\documentclass[a4paper,conference,final]{IEEEtran}

\usepackage{cite}
\ifCLASSINFOpdf
   \usepackage[pdftex]{graphicx}
\else
   \usepackage[dvips]{graphicx}
\fi

\usepackage[cmex10]{amsmath}
\usepackage{bbm}

\usepackage{clrscode}
\usepackage{algorithmic}
\usepackage{algorithm}

\usepackage{enumerate}

\usepackage{eqparbox}

\usepackage[caption=false]{subfig}

\usepackage{stfloats}

\usepackage{url}

\usepackage{color}
\usepackage{amssymb}
\usepackage{calc}
\usepackage{booktabs}

\newcommand{\refFig}[1]{Figure \ref{#1}}
\newcommand{\refSec}[1]{Section \ref{#1}}

\definecolor{algComment}{rgb}{0.5,0.5,0.5}

\usepackage{lipsum}
\usepackage{blindtext}
\usepackage[english]{babel}

\usepackage{flushend}
\usepackage{microtype}
\usepackage[printonlyused]{acronym}

\usepackage{mdwlist}

\usepackage[textsize=small, colorinlistoftodos]{todonotes}

\definecolor{blind-gray}{gray}{0.9}

\begin{document}

%\IEEEoverridecommandlockouts
%\IEEEpubid{\makebox[\columnwidth]{978-1-4673-2480-9/13/\$31.00 ~\copyright~2013 IEEE \hfill} \hspace{\columnsep}\makebox[\columnwidth]{ }}

\title{Dynamic Backhaul Network Configuration in SDN-based Cloud RANs}

\author{
\IEEEauthorblockN{Martin Dr\"axler}
\IEEEauthorblockA{University of Paderborn\\Paderborn, Germany\\
E-mail: martin.draexler@upb.de}
\and
\IEEEauthorblockN{Holger Karl}
\IEEEauthorblockA{University of Paderborn\\Paderborn, Germany\\
E-mail: holger.karl@upb.de}}

\maketitle

\begin{abstract}
The coordination of base stations in mobile access networks is an important approach to reduce harmful interference and to deliver high data rates to the users. Such coordination mechanisms, like Coordinated Multi-Point (CoMP) where multiple BSs transmit data to a user equipment, can be easily implemented when centralizing the data processing of the base stations, known as Cloud RAN.

This centralization also imposes significant requirements on the backhaul network for high capacities and low latencies for the connections to the base stations. These requirements can be mitigated by (a) a flexible placement of the base station data processing functionality and by (b) dynamically assigning backhaul network resources.
We show how these two techniques increase the feasibility of base station coordination in dense mobile access networks by using a heuristic algorithm.

We furthermore present a prototype implementation of our approach based on software defined networking (SDN) with OpenDaylight and Maxinet.
\end{abstract}

% \begin{IEEEkeywords}
% Keywords...
% \end{IEEEkeywords}

\IEEEpeerreviewmaketitle

\section{Introduction}
\label{sec:intro}
To handle the increasing wireless data rate demands of users in 5G mobile networks \cite{CiscoVNImobile}, it is not only important to efficiently utilize the available wireless resources to provide high data rates but also to coordinate wireless transmissions to reduce harmful interference.
Such coordination mechanisms require a set of base stations (BSs) that implement a coordination scheme to serve one or a group of user equipments (UEs). We call this set of BSs a \emph{coordinated base station set (CBS)}. Hence, as the coordination mechanisms are implemented in the wireless domain, the selection of BSs for the CBSs has to be decided based on the characteristics of the wireless channel\cite{6472201,6328484}. Implementing coordination among BSs is only possible if capacity and low-latency connections are provided by the backhaul network that 
interconnects the BSs. Thus, the constraints on the backhaul network have to be considered in addition to the wireless characteristics when implementing BS coordination mechanisms \cite{6649363,Biermann2011a,Biermann2012662}.

In our previous work \cite{draexlerew2014}, we have shown how dynamic backhaul reconfiguration can increase the feasibility of base station coordination in networks with limited backhaul network resources, but we did not take the new challenges from very dense wireless access networks into account. Thus, in this paper we extend our dynamic backhaul reconfiguration approach to handle hotspot areas with a high density of users, i.e. a high density of desired CBS. We also show that this extension is beneficial for scenarios without hotspots and how it dynamically adapts to the density of hotspots.

We furthermore present a prototype implementation of our approach that is based on OpenDaylight \cite{odl}, a production-grade SDN controller, and Maxinet \cite{maxinet}, a distributed emulation framework for OpenFlow SDNs.

Base station coordination can be implemented in different ways. A very promising implementation strategy is to include the coordination into the baseband unit (BBU) of a Cloud RAN. Existing work \cite{chinacran,6736747,ijoincommag,6923535,6897914} has already shown that using Cloud-RAN is a good option to implement in future dense wireless access networks, in particular in conjunction with the SDN paradigm \cite{6845049,CrowdEWSDN,onfsdn}.

Another approach to base station coordination is the idea to apply the software defined networking (SDN) paradigm also to the control and coordination of BSs. There the control of coordination mechanisms is also moved to a local controller (LC) node, thus constraints similar to CoMP between the controller node and the BSs also apply here. Different approaches for that have been proposed \cite{CrowdEWSDN,Bansal2012}.

The required flexibility and capacity in the backhaul network for our approach can be provided by optical backhaul networks based on WDM-PON.
WDM-PON \cite{RamaswamiOptical10} is an optical fiber technology that multiplexes optical carrier signals on an optical fiber using different wavelengths or colors of laser light.
This allows great flexibility in deploying and operating backhaul networks: wavelengths can be added, dropped or manipulated in network 
nodes, allowing to build dynamic topologies in the backhaul network. This flexibility is mainly enabled by the fact that
WDM allows operations like multiplexing different wavelengths and converting wavelengths in a purely optical way, avoiding slow and energy-consuming optical-electrical-optical conversion. Using optical links in a Cloud RAN backhaul has also been recommended by existing work \cite{6897914,6692220}.

The rest of the paper is structured as follows: We first discuss related approaches in \refSec{sec:relwork}. Then we describe our extended algorithm in \refSec{sec:algo} and our prototype implementation in \refSec{sec:prototype} including an overall system architecture. In \refSec{sec:eval} we provide both simulation results on the performance of our approach and on the feasibility of our prototype implementation. Finally, we conclude our work in \refSec{sec:concl}.

\section{Related Work}
\label{sec:relwork}
Bartelt et al. \cite{6692220} analyze the challenges for backhaul networks arising from a cloud-based RAN and conclude that optical networks are a promising technology for Cloud RAN backhauls, if the data rate and latency demands are included into a joint optimization of radio access and backhaul.
The challenges of the backhaul connection of small cells in heterogeneous mobile access networks is studied in \cite{6666414}, resulting in a novel CBS selection algorithm to include the backhaul constraints.
Liu et al. \cite{6566903} investigate a dynamic backhaul network for the dynamic assignment of base stations to BBU pools with a reconfigurable backhaul architecture, including a small testbed evaluation. 

The issues in backhaul networks with limited resources together with implementing wireless coordination have been investigated in in existing work: 
Evaluations from Biermann et al. \cite{Biermann2011a} show how a limited backhaul network reduces the efficiency of CoMP. 
Soliman et al. \cite{6666286} analyze how the backhaul resources have to be shared to achieve a feasible data exchange between coordinated BSs, but their model only considers two BSs.
The effects of a constrained backhaul on different BSs coordination schemes are investigated in \cite{6213935}, but without considering the selection of feasible CBSs.

\section{Heuristic Algorithm}
\label{sec:algo}
To deploy our algorithm in dense scenarios with hotspots of CBSs, we extend the heuristic algorithm from our previous work \cite{draexlerew2014} by a CBS prioritization mechanism.
This mechanism is explained in detail in this section together with a brief explanation of the whole algorithm.

The inputs for the heuristic are the backhaul network as an annotated graph $G = (V,E)$, a set of available wavelengths per link $K$ and the desired CBS $W$ with each $W_i \subset V$ together with their constraints on data rate and latency.

An overview of the heuristic is depicted in \refFig{tab:bfsclust} and the individual steps are explained below.

% \floatstyle{boxed} 
% \restylefloat{figure}
\begin{figure}[tb]
%\tdi{bild waere schoener}
%\begin{minipage}{0.45\textwidth}
\fbox{%
\parbox{0.469\textwidth}{%
\small
%\begin{enumerate}
%\item CBS Prioritization\\
\textbf{CBS Prioritization}\\
$\rhd$ decide for each CBS if it belongs to a hotspot\\
$\rhd$ \emph{output}: list of hotspot CBSs and normal CBSs
%\end{enumerate}

\textbf{For each CBS in $\text{W}_\text{hotspot}$, then for each CBS in $\text{W}_\text{normal}$}
\begin{enumerate}
\item \textbf{Maximum-Path BFS}\\
% $\rhd$ start modified BFS from all vertices\\
% $\rhd$ \emph{output}: BFS trees for all vertices
$\rhd$ start modified BFS from each vertex\\
$\rhd$ \emph{output}: BFS tree for each vertex
\item \textbf{Match CBS}\\
$\rhd$ match BFS trees against CBS\\
$\rhd$ \emph{output}: possible candidate BFS trees for CBS
\item \textbf{Back-Track BFS Trees}
\begin{enumerate}
\item \textbf{Check constraints}\\
$\rhd$ recheck constraints on candidate BFS trees
\item \textbf{Wavelength Assignment}\\
$\rhd$ determine wavelengths for all candidate BFS trees
\end{enumerate}
$\rhd$ \emph{output}: possible candidate BFS trees and their wavelength assignment for CBS
% \item Back-Track BFS Trees\\
% $\rhd$ recheck constraints on candidate BFS trees\\
% $\rhd$ \emph{output}: possible candidate BFS trees for CBS
% \item Wavelength Assignment\\
% $\rhd$ determine wavelengths for all candidate BFS trees\\
% $\rhd$ \emph{output}: wavelength assignment for all possible candidate BFS trees for CBS
\item \textbf{Match CBS}\\
$\rhd$ match BFS trees against CBS again\\
$\rhd$ \emph{output}: confirmed candidate BFS trees for CBS
\item \textbf{Find Best BFS Tree}\\
$\rhd$ compare candidate BFS trees\\
$\rhd$ \emph{output}: best BFS tree for CBS
\end{enumerate}
}
}
%\end{minipage}
\caption{Heuristic algorithm overview}
\label{tab:bfsclust}
\end{figure}
% \floatstyle{plain}
% \restylefloat{figure}

\subsubsection*{CBS Prioritization}
This new step of the algorithm divides the desired CBS into two sets $\text{W}_\text{hotspot}$ and $\text{W}_\text{normal}$. 
The calculations in this step use two threshold values $t_v$, which determines how strict the filtering of hotspot vertices should be, and $t_h$, which determines how strict the filtering of hotspot CBS should be. The numbers for the threshold values depend on the scenario and can be determined using a parameter study.

The algorithm first calculates for each vertex $v$ from the backhaul graph in how many CBS the vertex is present as:
% \begin{align*}
% h_v = \sum_{W_i \in W}{
% \begin{cases}
% v \in W_i & 1\\
% \text{else} & 0
% \end{cases}}
% \end{align*}
\begin{align*}
h_v = \mathbbm{1}_{v \in W_i}
\end{align*}
$\mathbbm{1}_{X}$ denotes an indicator variable with value $1$ if condition $X$ is true and value $0$ otherwise.
Each vertex $v$ is considered a hotspot vertex if
\begin{align*}
%\frac{\sum_{v \in V}{h_v}}{h_v} < |V| \cdot t_v
h_v > |W| \cdot t_v
\end{align*}
and is added to the set of hotspot vertices $V_h$. Now each CBS $W_i$ is added to $\text{W}_\text{hotspot}$ if
% \begin{align*}
% \frac{\sum_{v \in W_i}{
% \begin{cases}
% v \in V_h & 1\\
% \text{else} & 0
% \end{cases}
% }}{|W_i|} \geq t_w\texttt{}
% \end{align*}
\begin{align*}
\frac{\sum_{v \in W_i} \mathbbm{1}_{v \in V_h}}{|W_i|} \geq t_h
\end{align*}
otherwise it is added to $\text{W}_\text{normal}$. 

Now the actual algorithm for BBU/LC placement and backhaul resource allocation has to be executed for each CBS. The following steps are executed per CBS, starting with the CBSs from $\text{W}_\text{hotspot}$ and the CBSs from $\text{W}_\text{normal}$:

\subsubsection{Maximum-Path BFS}
The heuristic performs a modified breadth-first search from each vertex of the graph as the start node. Whenever a new tree edge $(u,v)$ is discovered, the constraints for the new edge are checked in the following way:
\begin{itemize}
\item Does the latency to the new vertex $v$ via the whole path from the root via $u$ and the new edge $(u,v)$ not exceed the maximum round-trip latency?
\item Are there any wavelengths on $(u,v)$ with enough free capacity to connect $v$ to the root vertex, if $v$ is part of the CBS?
\end{itemize}
If these checks are not successful, the new vertex $v$ is discarded, otherwise it is added to the tree.
The result of this step is a set $T$ of BFS trees, which only contain vertices that meet the constraints within the tree. This does not yet take into account reciprocal effects between multiple BSs in the CBS.

\subsubsection{Match CBS}
The heuristic checks for every BFS tree $T_i$ from step \emph{2)}, whether it contains all BSs from the desired CBS. The result is a set of only these BFS trees $T$ that match the desired CBS.

\subsubsection{Back-Track BFS Trees}
%The idea of this step is quite similar to the modified BFS in step \emph{1)}. 
In the constraint checking in step \emph{1)} vertices were discarded based on the latency for the full paths to the start
node and the link capacity for only single links and not the full paths.
Here the capacity constraints are checked again taking into account the capacities on full paths.

If the constraints are not violated a wavelength assignment for all routing paths between the BSs from the CBS and the BFS tree root has to be determined. Because the required data structures for this step can easily be constructed in the back-tracking phase, the back-tracking and the wavelength assignment are executed in one step.

Details on the implementation of the wavelength assignment can be found in our previous work \cite{draexlerew2014}.

\subsubsection{Match CBS}
After removing vertices in the previous step, the data on trees matching the CBS is not valid anymore. Thus, step \emph{2)} is repeated. If a tree still matches the CBS, the starting node is a valid candidate for a controller location for that CBS.

\subsubsection{Find Best BFS Tree}
Finally, the best BFS tree from the remaining candidates has to be determined. The algorithm calculates three different costs per candidate tree and sums them up in a weighted sum $n$ per tree
\begin{align*}
n = \sum (w_g \cdot n_g + w_a \cdot n_a + w_l \cdot n_l)
\end{align*}
with
\begin{itemize*}
\item $n_g$ as the total number of wavelengths used in the tree
\item $n_a$ as the number of wavelengths that have to assigned additionally for using that tree
\item $n_l$ as the number of used links
\end{itemize*}
and $w_g$, $w_a$ and $w_l$ as weight factors.

The algorithm then selects the tree with the lowest total cost, sets the root BS as the controller, stores the routing paths and updates the annotations on $G$ for running the next iteration for the next CBS.

\section{Prototype}
\label{sec:prototype}
In this section we first describe the architecture of a real-world implementation of our approach and then introduce how we implemented our prototype.
\subsection{System Architecture}
\label{sec:system}

\begin{figure}[hbt]
%\begin{figure}[H]
  \begin{center}
  \includegraphics[width=0.5\linewidth]{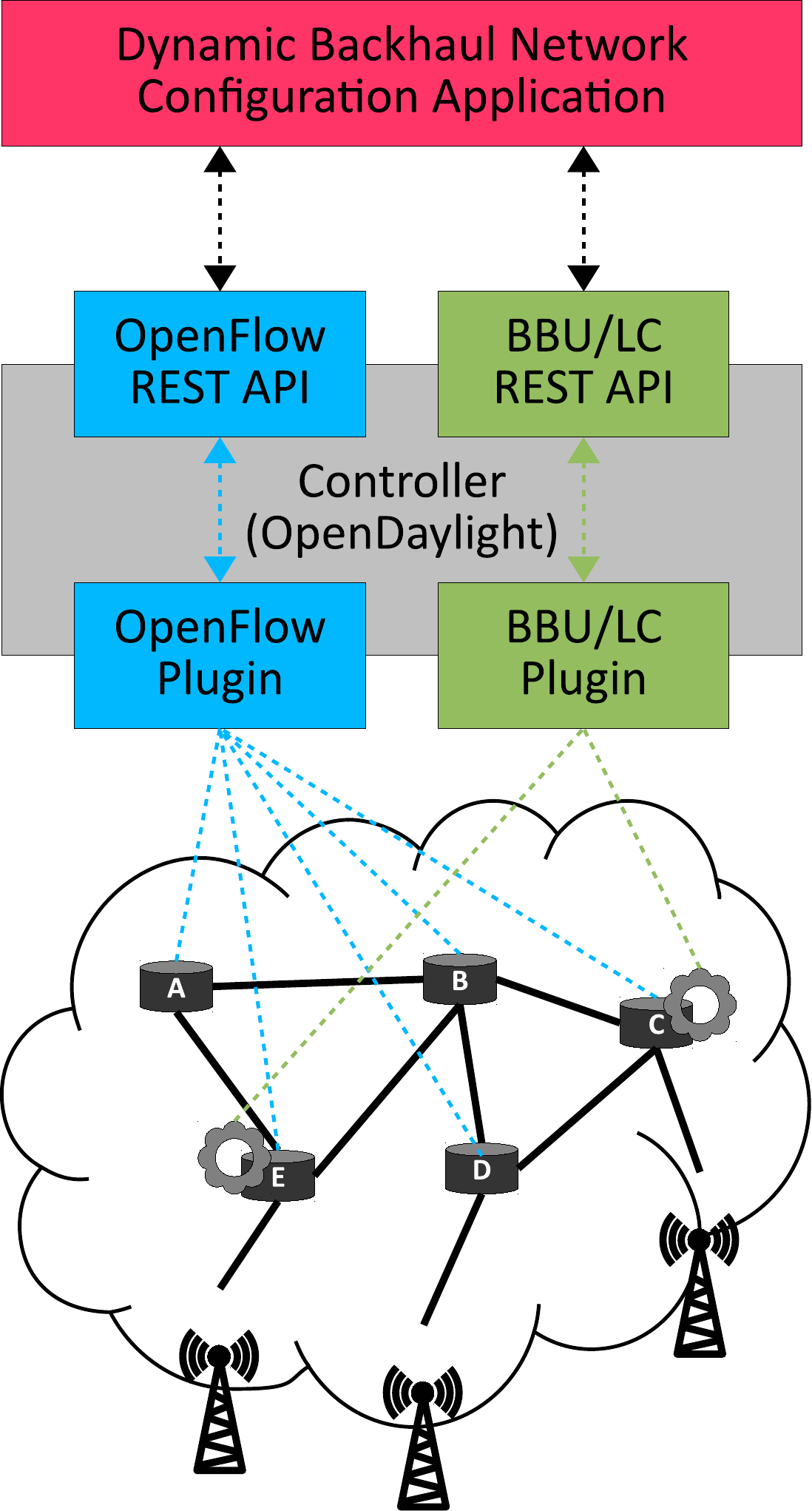}
  \caption{System architecture}
  \label{fig:architecture}
  \end{center}
\end{figure}

Our system architecture is based on the decoupling of the following three components of the overall system:
\begin{itemize*}
 \item Application
 \item Controller
 \item Network
\end{itemize*}
The architecture of our system is shown in \refFig{fig:architecture}.

The application is essentially the algorithm from \refSec{sec:algo} together with necessary data structures to store the input and output data at runtime. The application is decoupled from the controller by accessing two different APIs of the controller. With the OpenFlow API the application can both query the topology and flow configuration of the backhaul network and modify the flow configuration according to the algorithm outputs. In addition to this API, the application also needs to access the BBU/LC API in order to query the status of running BBUs/LCs and to start or stop BBUs/LCs.

The controller acts as the centralized link between the application and the backhaul network, like the controller in an SDN network \cite{onfsdn}. It exposes the previously described northbound APIs to the application and controls both the backhaul network and the BBU/LC instances via its southbound plugins. Consistent with the northbound APIs the controller needs both an OpenFlow plugin and a BBU/LC control plugin. 

Of course the backhaul network has to be based on OpenFlow-enabled hardware, i.e. switches, otherwise the controller could not use the OpenFlow plugin to reconfigure the backhaul network. Also all potential nodes for hosting a BBU/LC have to run a hypervisor to allow the dynamic instantiation of BBU/LC instances.

\subsection{Implementation}
\label{sec:impl}

Our reference implementation of the described system architecture is based on the OpenDaylight controller platform \cite{odl}. OpenDaylight already includes a fully featured OpenFlow plugin, thus no additional implementation is required for this part. Also the model-based design behind OpenDaylight facilitated the implementation of a BBU/LC control plugin. A core concept with OpenDaylight is the implementation of REST-based northbound APIs. The OpenFlow plugin already provides APIs for both querying the topology and configuring flows within the backhaul network. For our BBU/LC control plugin, we implemented REST APIs accordingly.

Our implementation of the application is based on Python and solely relies on the REST APIs of the controller. A small wrapper converts data between the OpenDaylight API format and the data structures for our algorithm. This wrapper is tailored to the OpenDaylight API format, but it could in principle be adapted to any other OpenFlow controller, making our algorithm implementation independent of the controller platform.

\subsection{Testbed}
\label{sec:testbed}

\begin{figure}[bt]
%\begin{figure}[H]
  \begin{center}
  \includegraphics[width=0.7\linewidth]{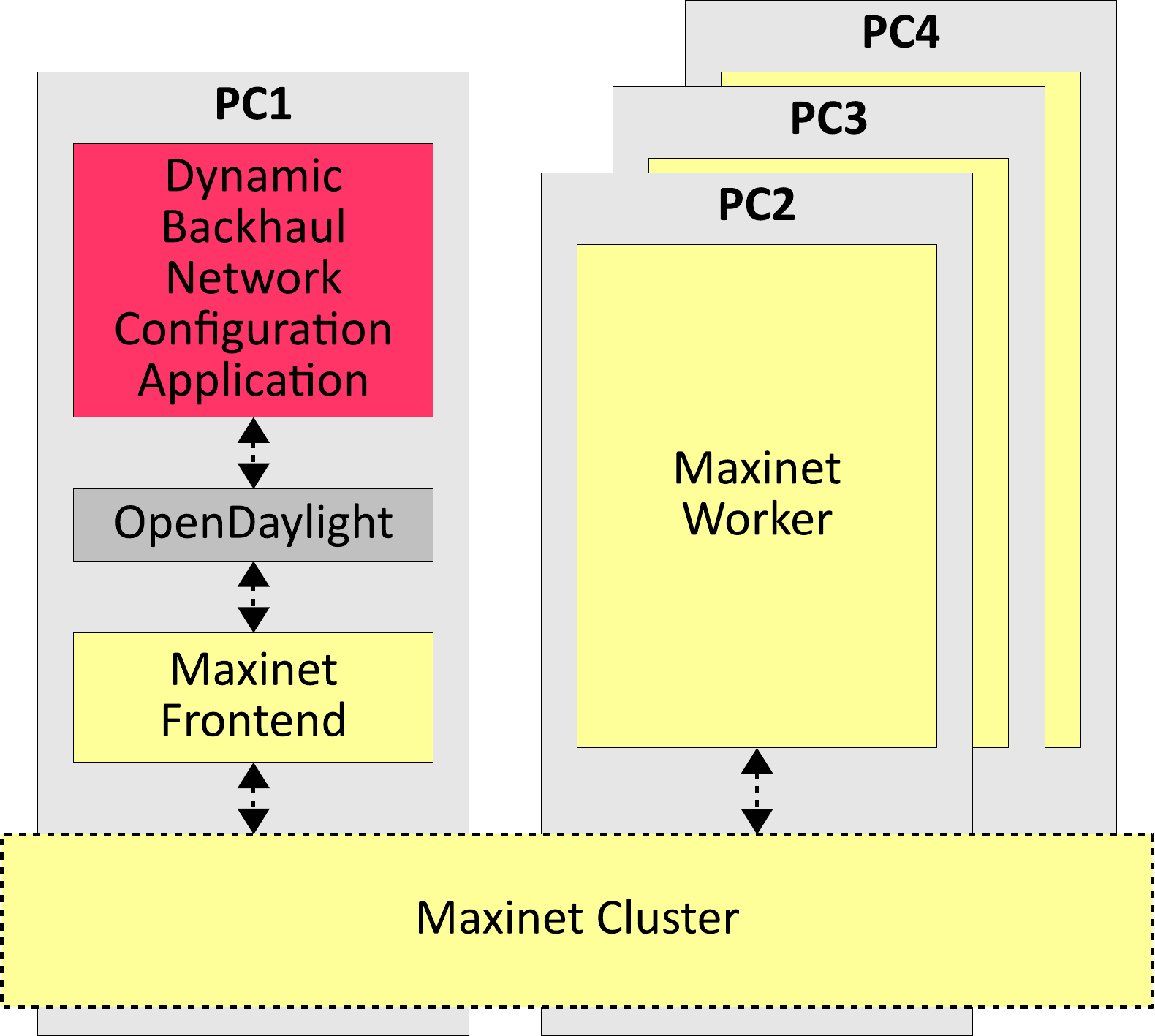}
  \caption{Testbed}
  \label{fig:testbed}
  \end{center}
\end{figure}

Because we cannot test our application and controller on a real-world OpenFlow enabled backhaul network, we have to use an emulated network to test and evaluate our implementation.
For this we use Maxinet \cite{maxinet}, an extension to the well known Mininet emulator \cite{mininet} for distributed emulation, to emulate a fully functional, virtual OpenFlow enabled network on a cluster of physical machines. We use a small setup with four physical machines, as shown in \refFig{fig:testbed}. One machine hosts the application, OpenDaylight and the Maxinet frontend, the three other machines are used as Maxinet workers to emulate the backhaul network.
Since we do not have any wireless interface integrated into the testbed, the traffic between the BSs and the BBUs/LCs is emulated using iperf \cite{iperf}.

\section{Evaluation}
\label{sec:eval}

In this section we present (a) simulation results that show how our new algorithm performs compared to the algorithm from our previous work \cite{draexlerew2014} in \refSec{sec:evalprio}. And (b) a testbed evaluation to show how our prototype implementation works in \refSec{sec:evalproto}.

\subsection{CBS Prioritization Simulation}
\label{sec:evalprio}

\subsubsection{Simulation Scenario}

A fixed number of BSs are placed on a regular grid, with a mean inter-BS distance of $\bar{s} = 1000\,\mathrm{m}$ (urban scenario \cite{6629714}), and are then shifted in both x and y direction according to two independent, normally distributed random variables with zero mean and standard deviation $\frac{\bar{s}}{8}$.

The backhaul network is created as a mesh topology where two BSs are connected by a link if their distance is less than or equal to $1.5 \cdot \bar{s}$. This value produces a partially connected mesh network; smaller or larger values result in too sparse or too dense topologies, which are not realistic.
All links in the backhaul network are assigned the same set of available wavelengths $K=4$ and each wavelength is assigned the same fixed capacity of 2.5\,Gb/s.
The latency for each link is determined by the distance multiplied by 1.45 divided by the speed of light as we are modeling an optical backhaul network.

In order to create a hotspot of CBSs, a x,y coordinate is selected uniformly as the hotspot center. Now a fraction $h$ of all CBSs is placed around the hotspot center based on a normal distribution with the hotspot coordinate as the mean and standard deviation $\frac{\bar{s}}{4}$ as the desired hotspot CBSs. All other desired CBSs are generated by placing then uniformly at random on the plane covered by the placed BSs. The BSs which are considered as the CBS are all BSs located inside a circle around the coordinates of the CBS with a given radius. We determine this radius by multiplying the mean inter-BS distance with a factor $r=1.5$, which results in 5 BSs per CBS on average.

The capacity demand $d$ for each BS in the CBS is set to the same value and is either 0.625\,Gb/s, 1.25\,Gb/s or 2.5\,Gb/s. This implies that at a demand of 2.5\,Gb/s one complete wavelength is required to connect a BS to the controller.

To determine the threshold values for the CBS prioritization $t_v$ and $t_h$, we performed a parameter study and identified $t_v=0.1$ and $t_h=0.9$ as the best values to maximize the number of feasible CBSs in this scenario.

%%% 3x4
\begin{figure*}[tbh]
  \begin{center}
    \subfloat[$h=0$]{
  \includegraphics[width=0.235\textwidth]{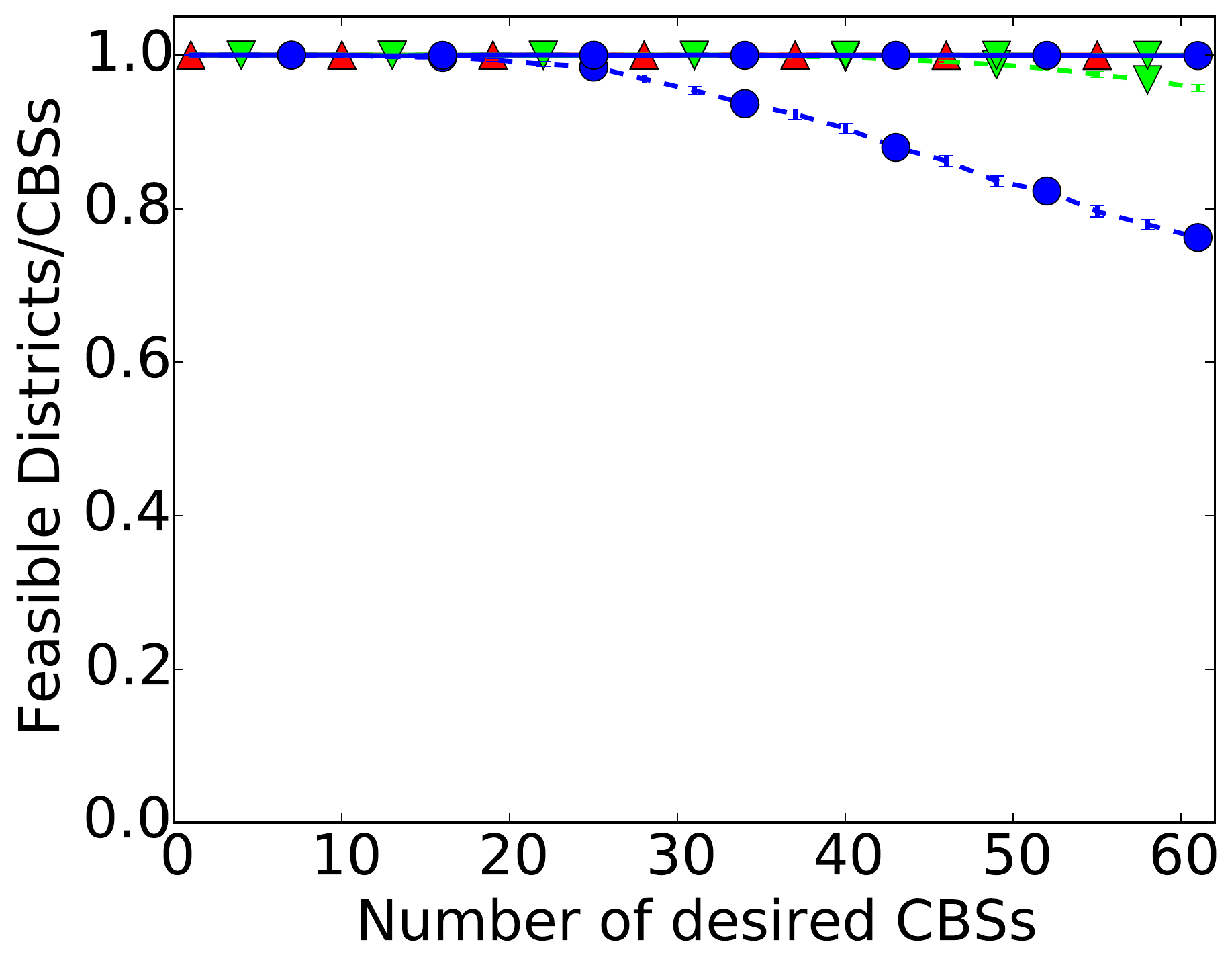}
  \label{fig:simfeas0}
  }%\\
%     \subfloat[$h=0.25$]{
%   \includegraphics[width=0.2\textwidth]{figures/sim_feas_025.pdf}
%   \label{fig:simfeas025}
%   }%\\
    \subfloat[$h=0.5$]{
  \includegraphics[width=0.235\textwidth]{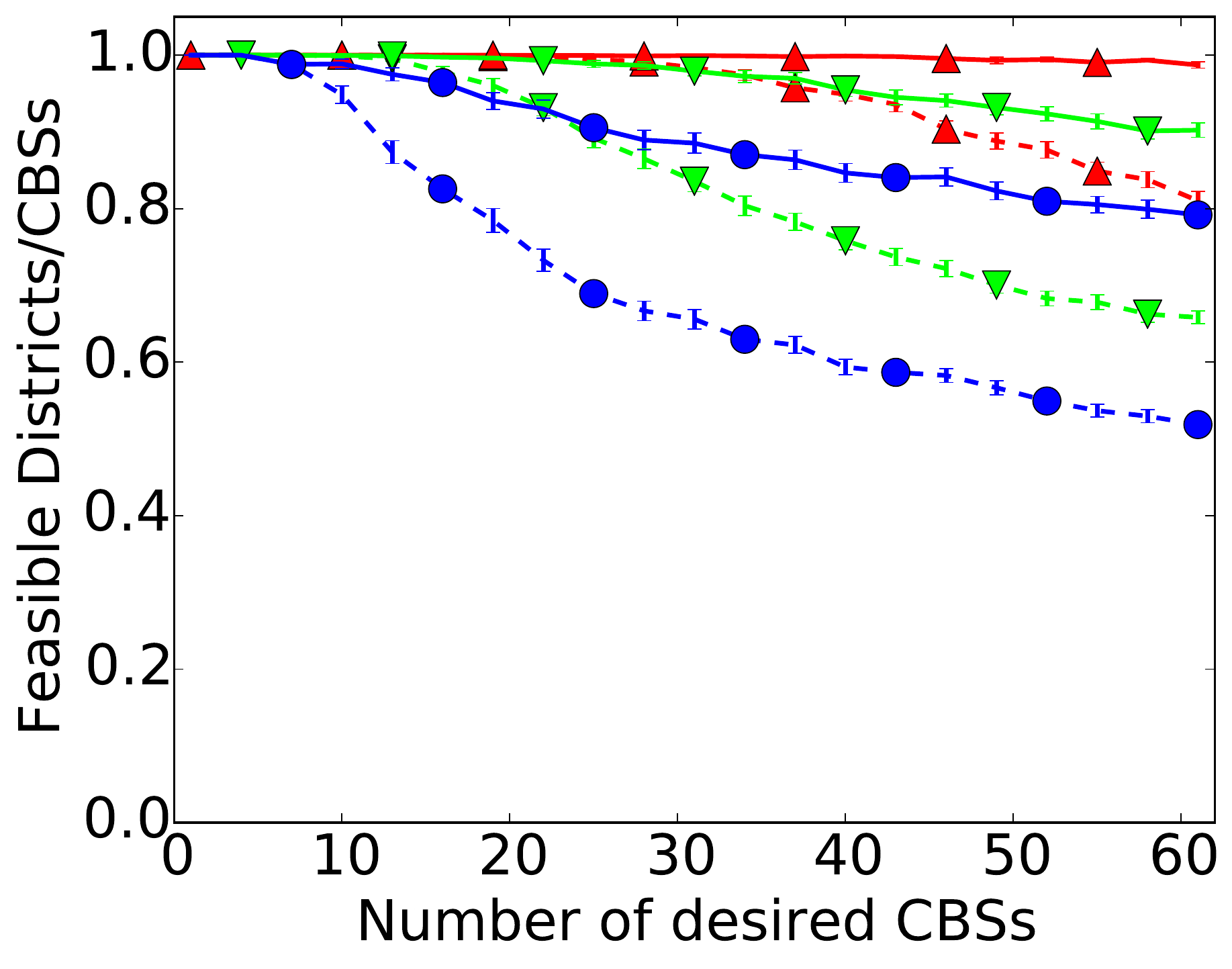}
  \label{fig:simfeas05}
  }%\\
    \subfloat[$h=0.75$]{
  \includegraphics[width=0.235\textwidth]{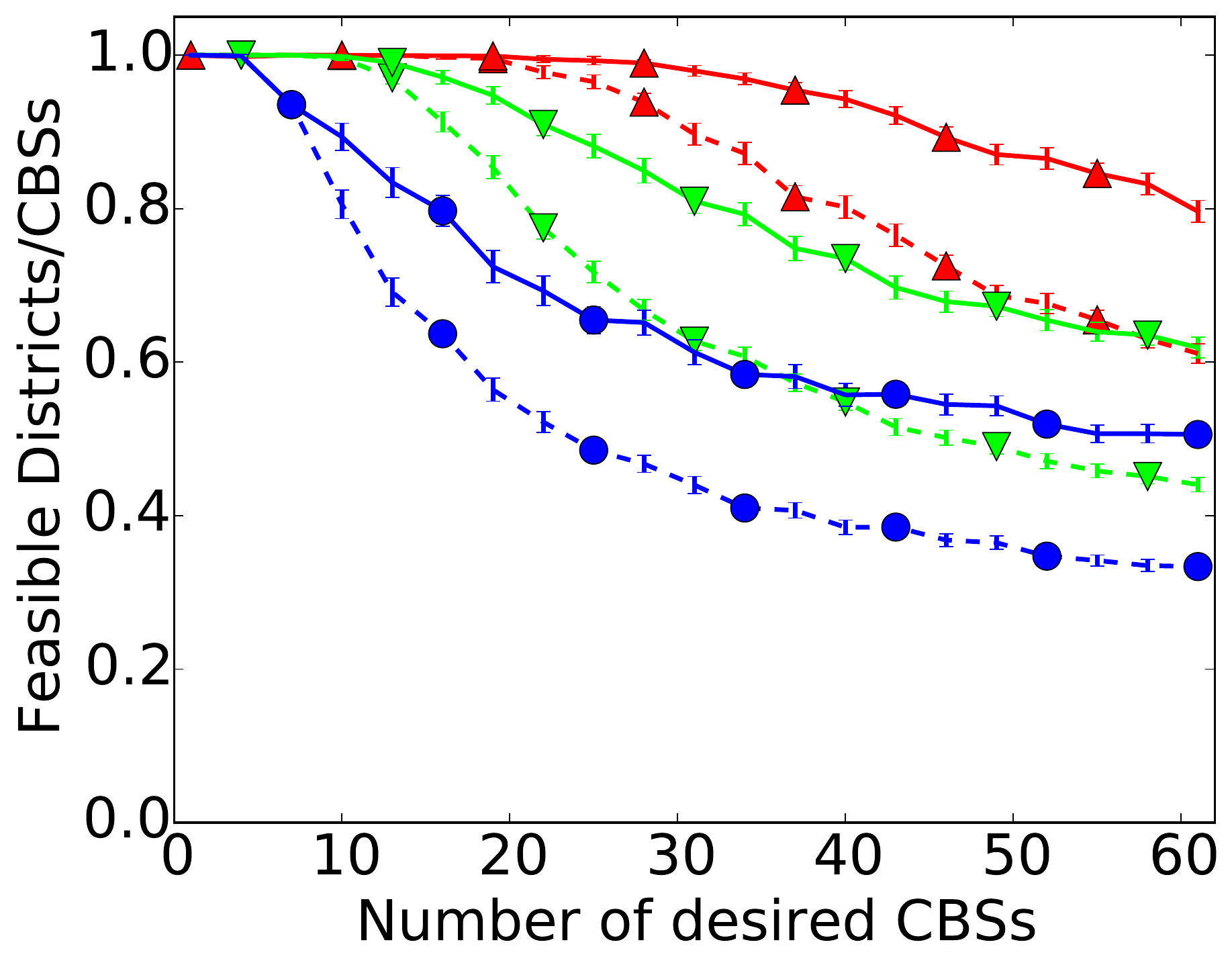}
  \label{fig:simfeas075}
  }%\\
      \subfloat[$h=1$]{
  \includegraphics[width=0.235\textwidth]{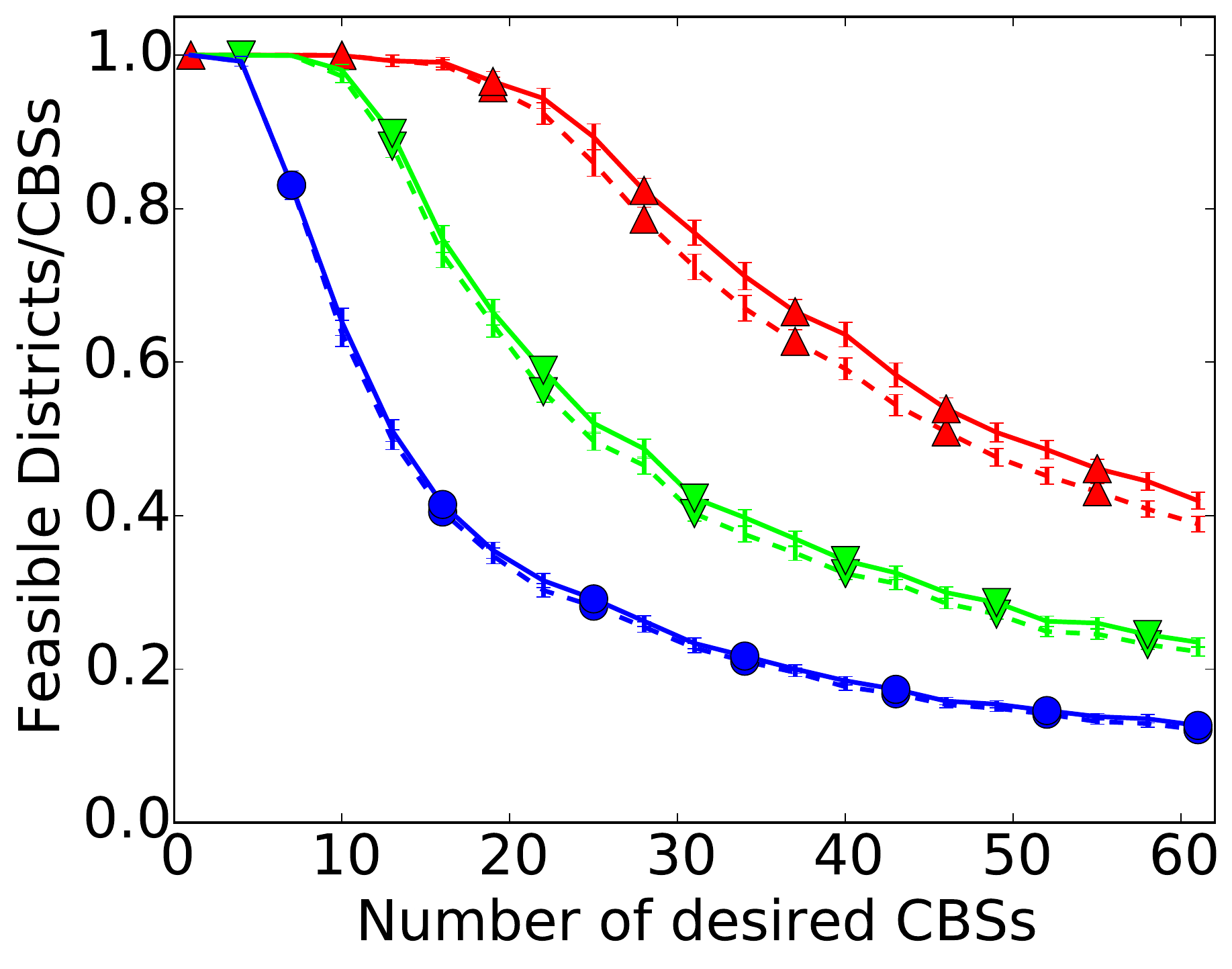}
  \label{fig:simfeas1}
  }
  \\
      \subfloat[$h=0$]{
  \includegraphics[width=0.235\textwidth]{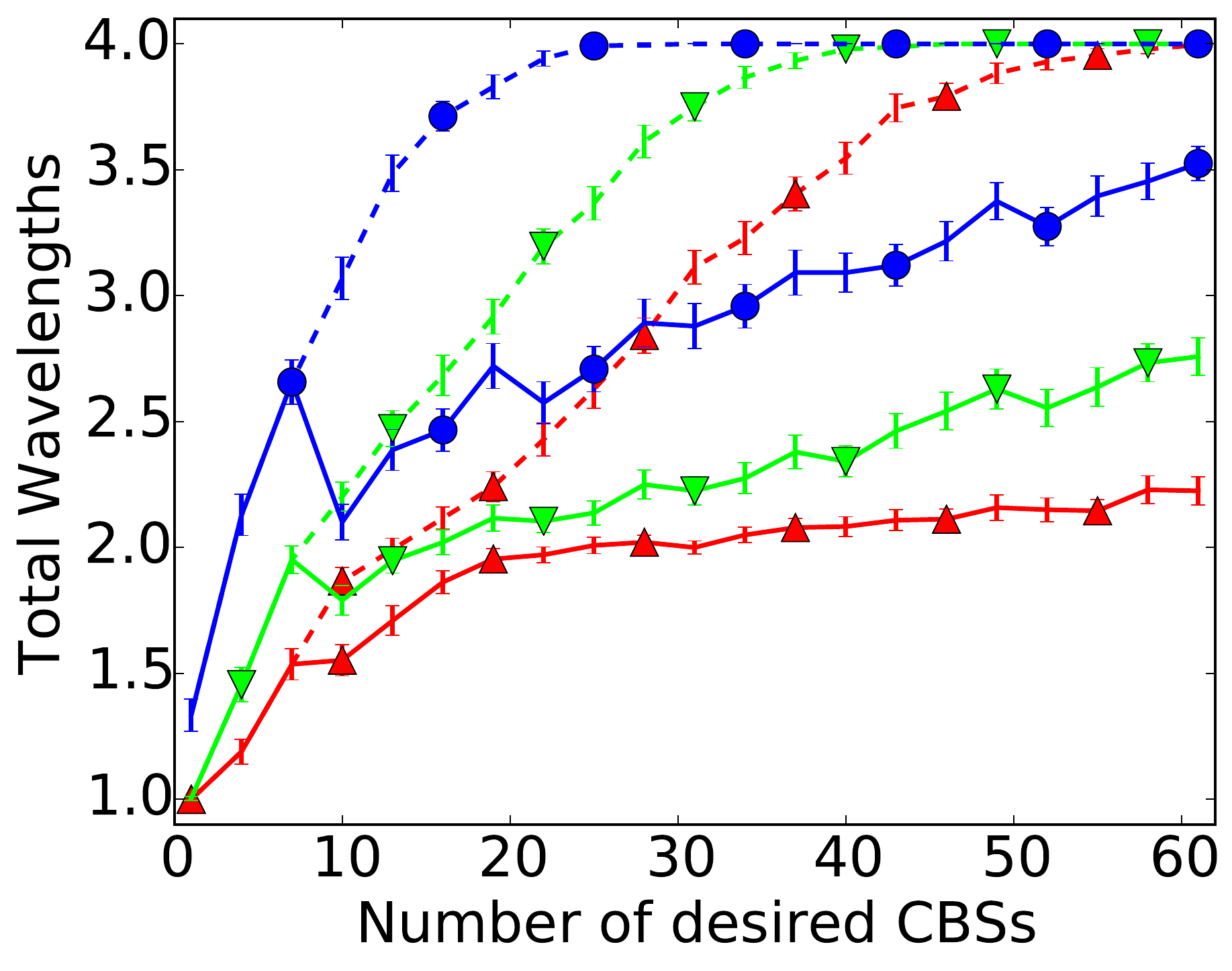}
  \label{fig:simwl0}
  }%\\
%     \subfloat[$h=0.25$]{
%   \includegraphics[width=0.2\textwidth]{figures/sim_wl_025.pdf}
%   \label{fig:simwl025}
%   }%\\
    \subfloat[$h=0.5$]{
  \includegraphics[width=0.235\textwidth]{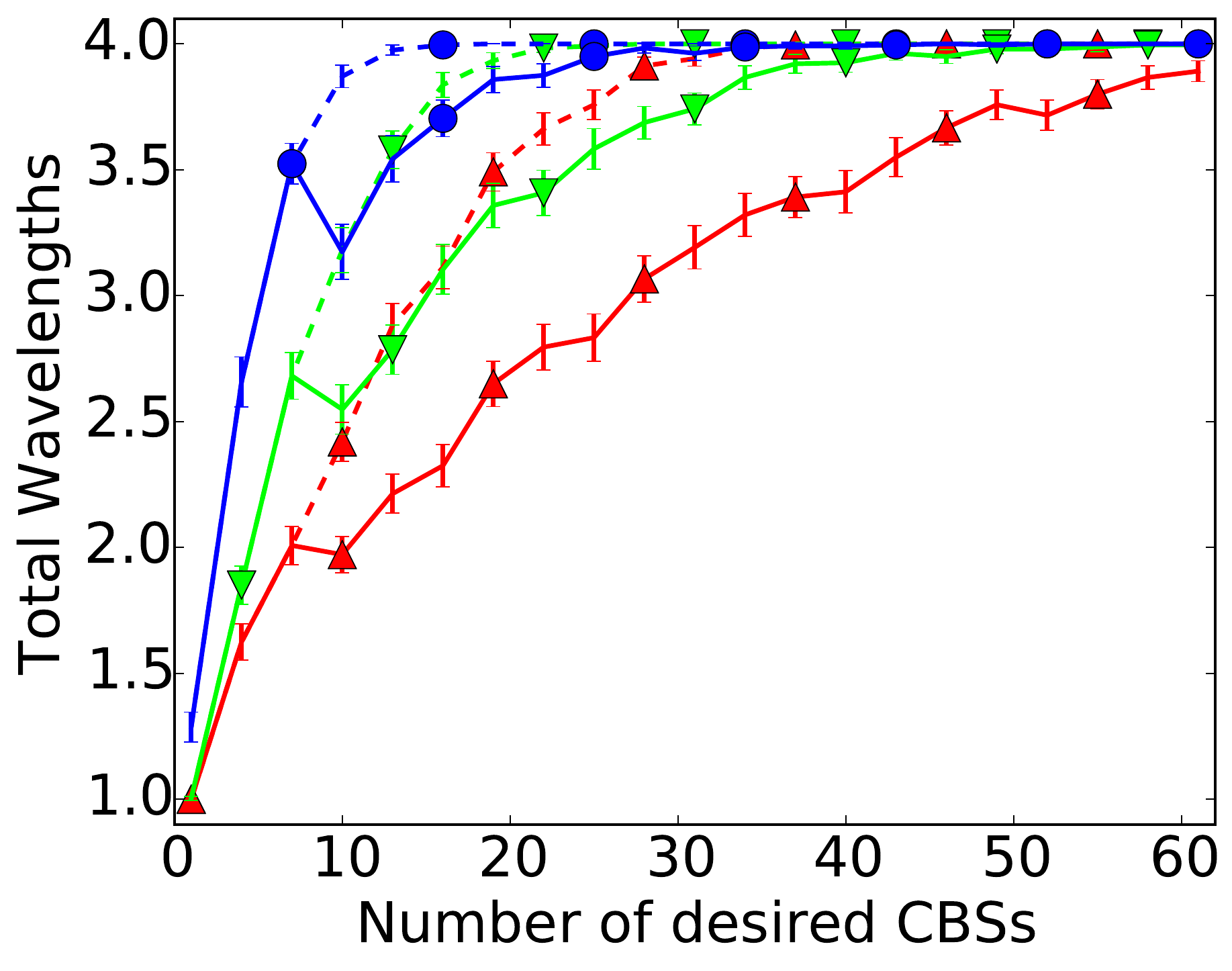}
  \label{fig:simwl05}
  }%\\
    \subfloat[$h=0.75$]{
  \includegraphics[width=0.235\textwidth]{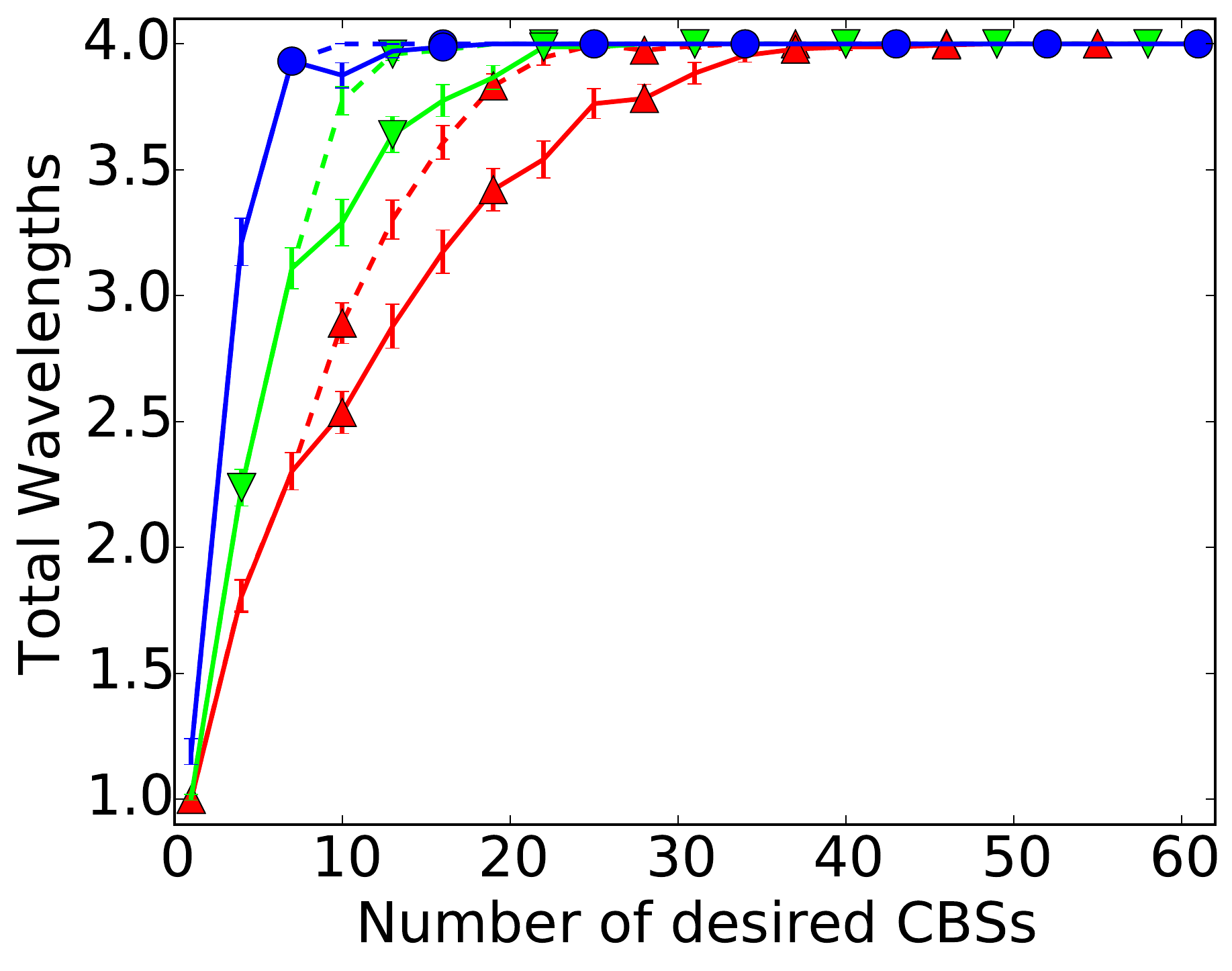}
  \label{fig:simwl075}
  }%\\
      \subfloat[$h=1$]{
  \includegraphics[width=0.235\textwidth]{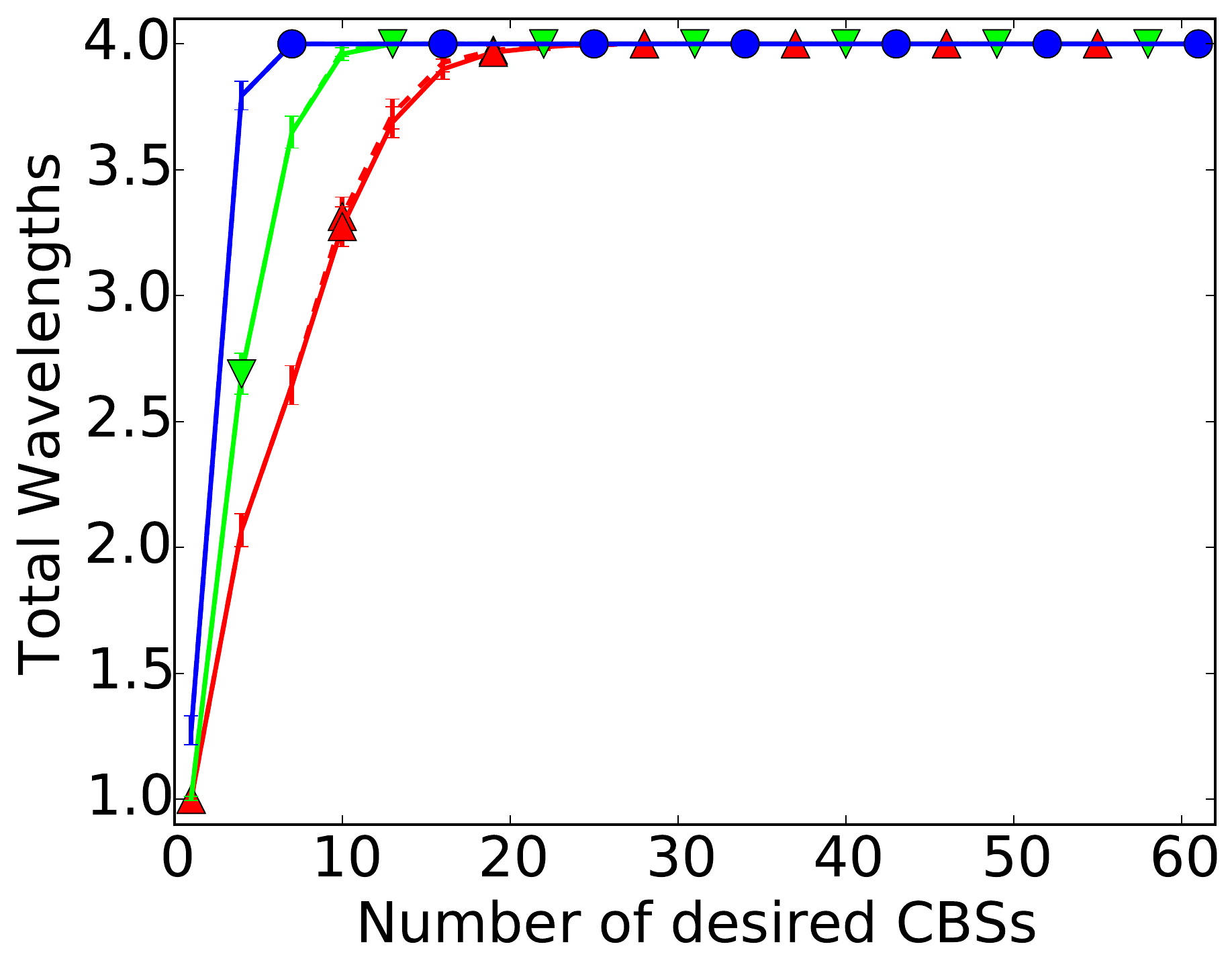}
  \label{fig:simwl1}
  }
    \\
      \subfloat[$h=0$]{
  \includegraphics[width=0.235\textwidth]{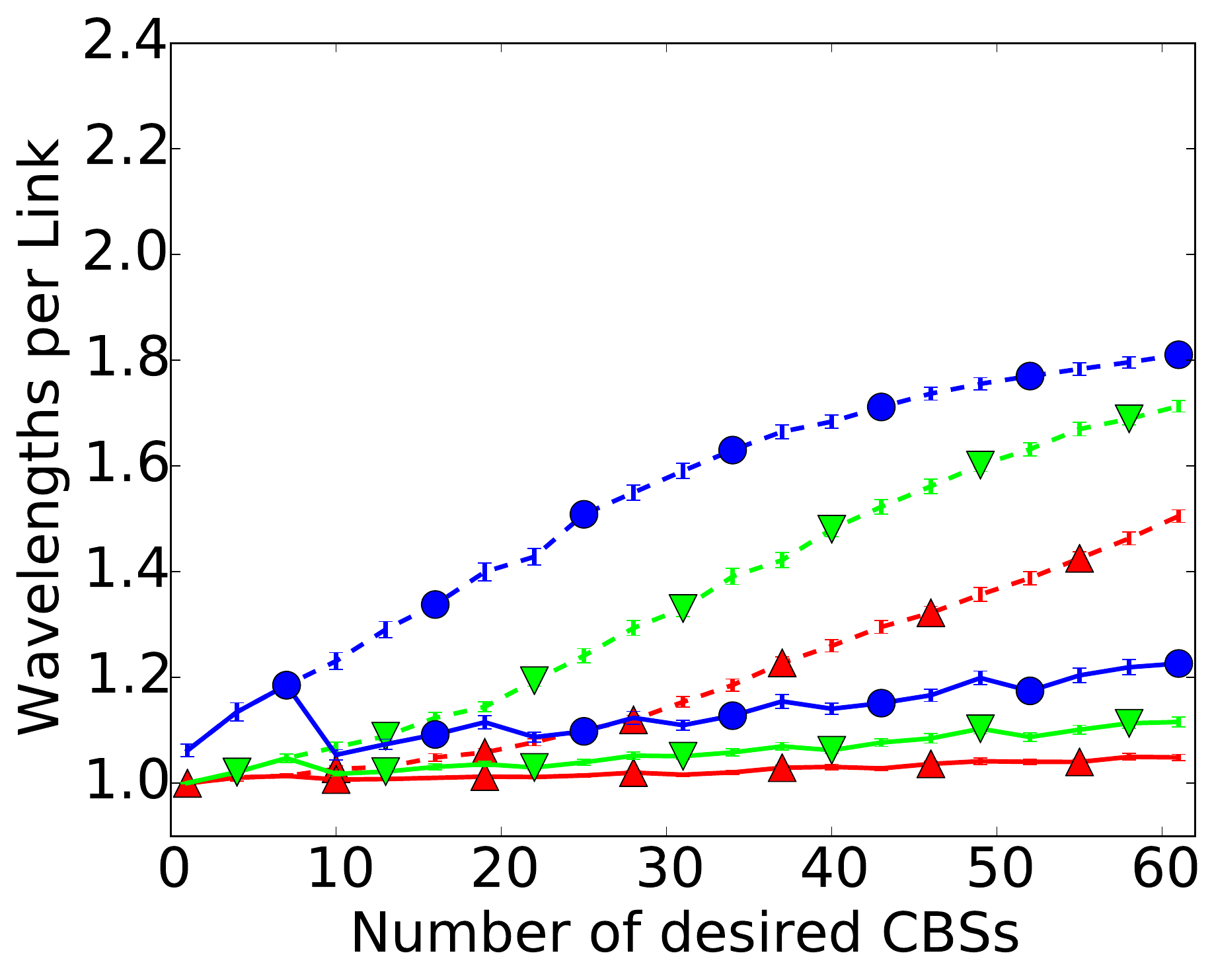}
  \label{fig:simwpl0}
  }%\\
%     \subfloat[$h=0.25$]{
%   \includegraphics[width=0.2\textwidth]{figures/sim_wpl_025.pdf}
%   \label{fig:simwpl025}
%   }%\\
    \subfloat[$h=0.5$]{
  \includegraphics[width=0.235\textwidth]{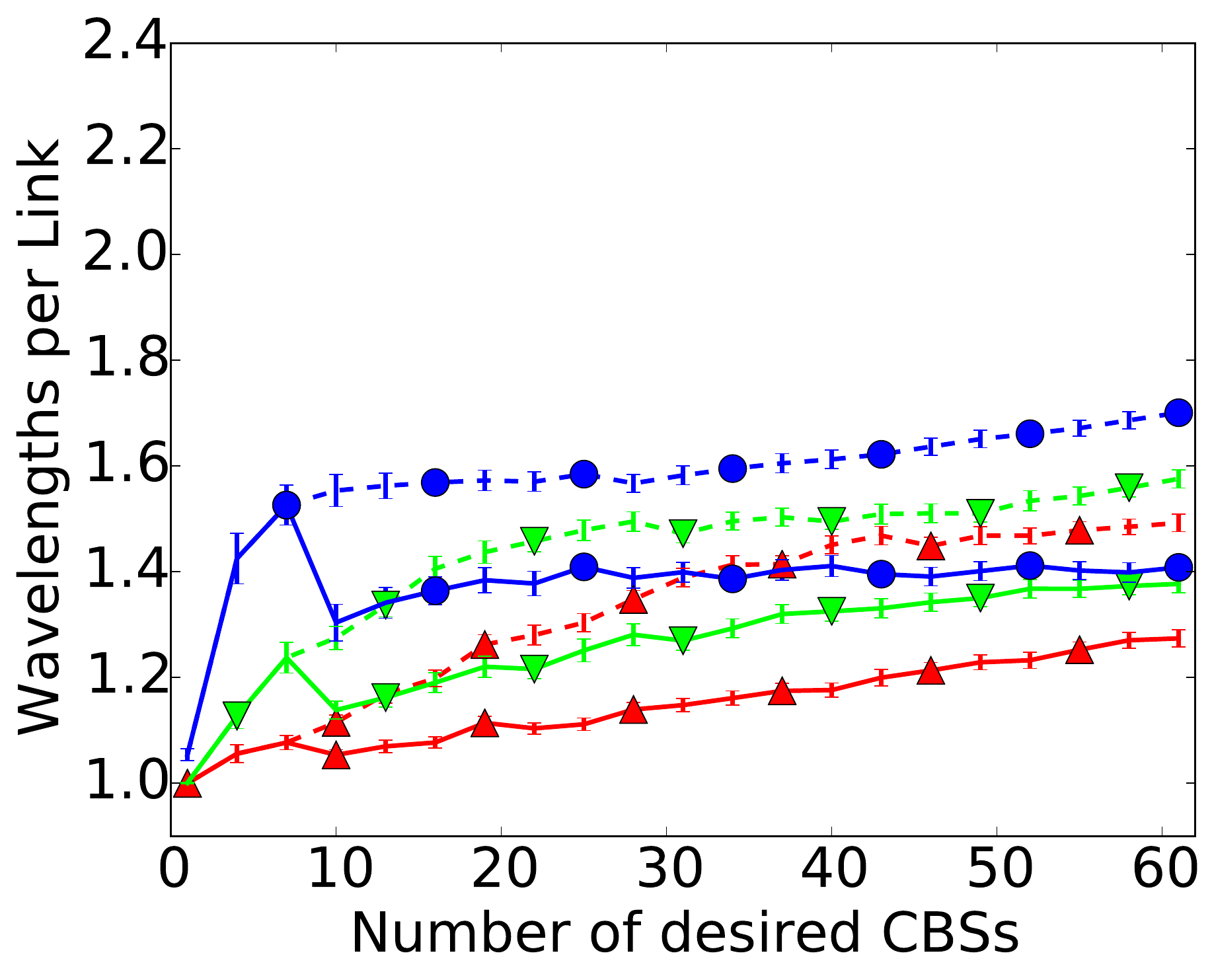}
  \label{fig:simwpl05}
  }%\\
    \subfloat[$h=0.75$]{
  \includegraphics[width=0.235\textwidth]{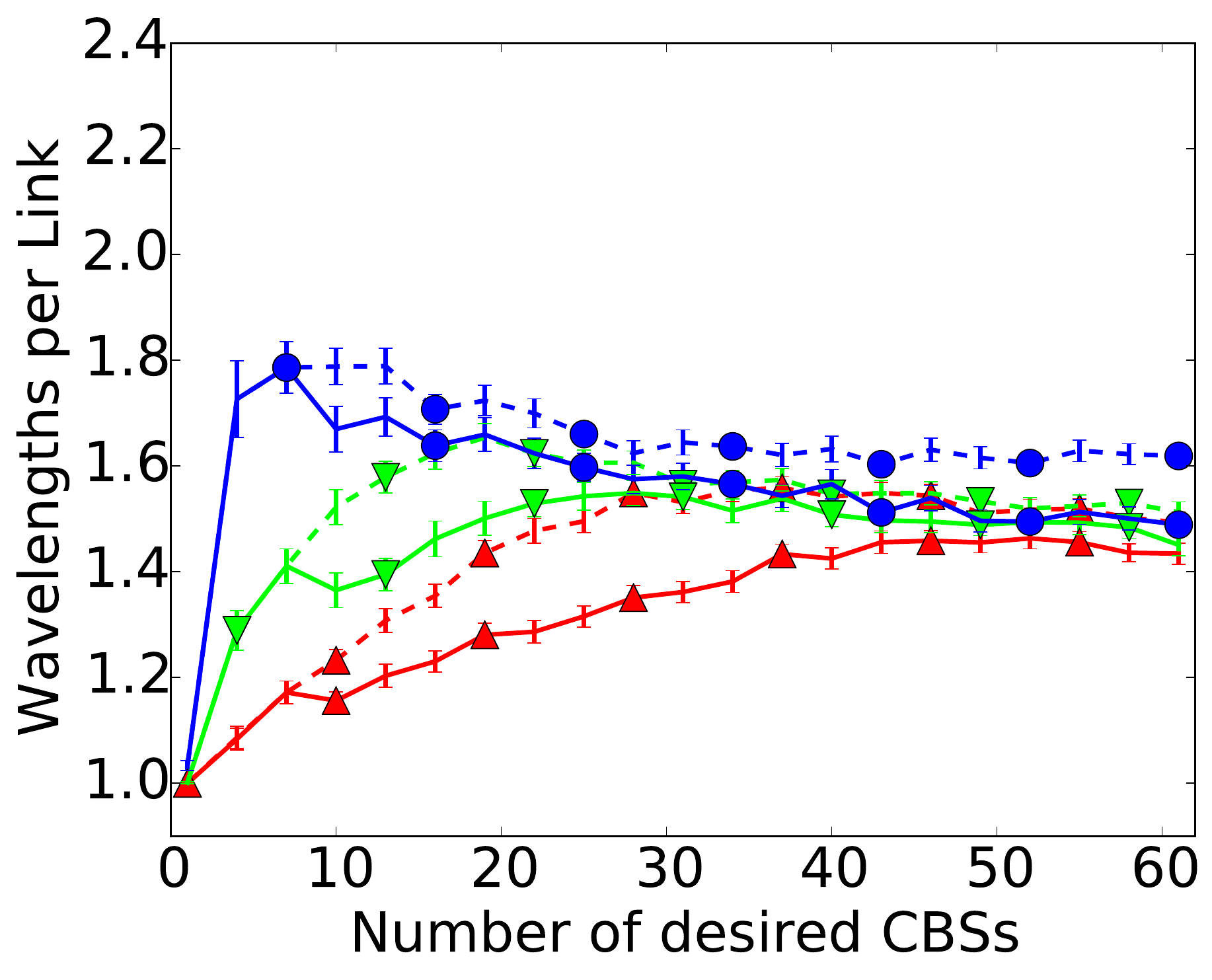}
  \label{fig:simwpl075}
  }%\\
      \subfloat[$h=1$]{
  \includegraphics[width=0.235\textwidth]{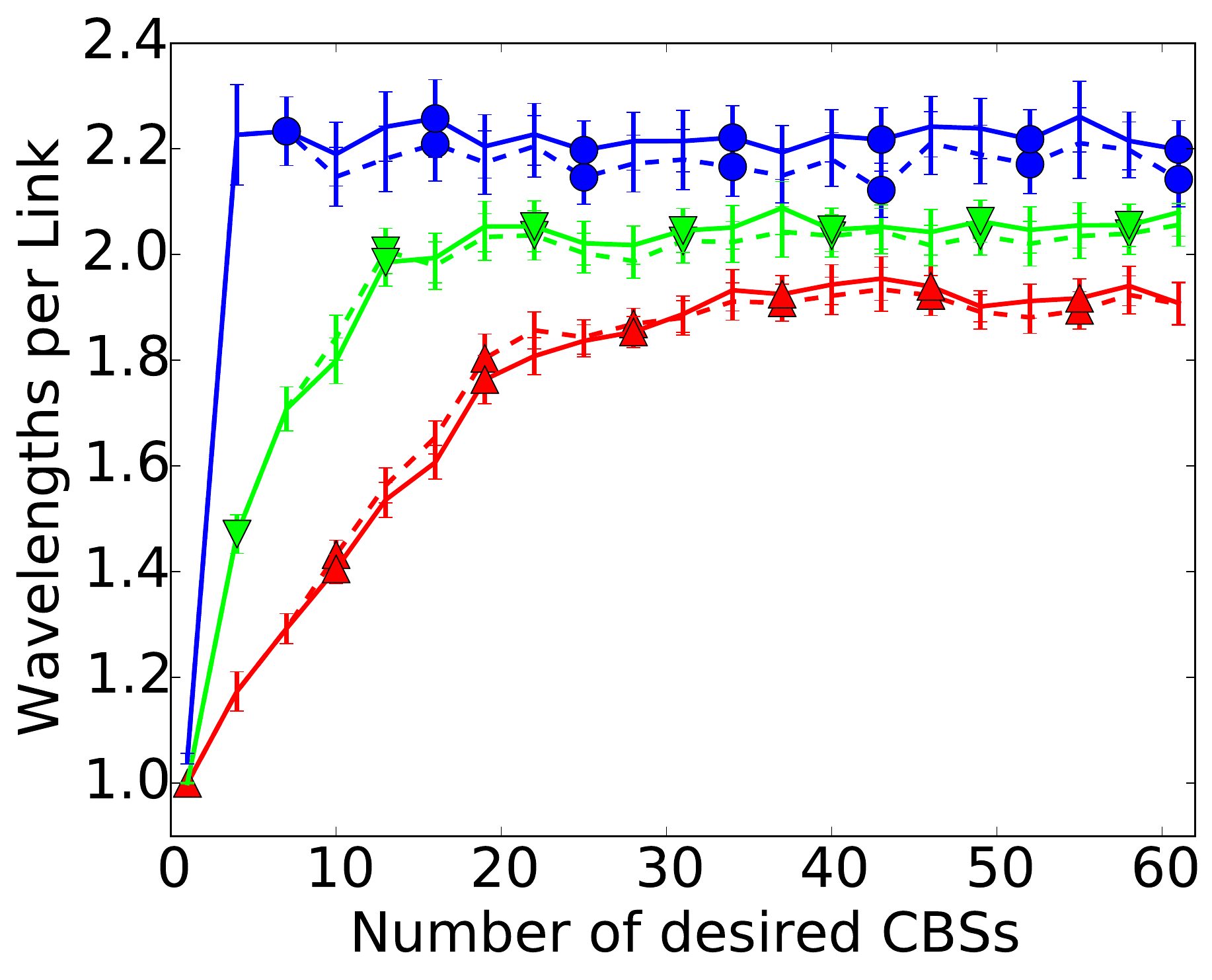}
  \label{fig:simwpl1}
  }
  \\
  \subfloat{
  \includegraphics[width=0.8\textwidth]{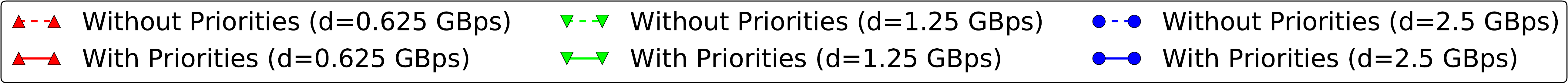}
  \label{fig:simlegend}
  }
  \caption{CBS prioritization simulation}
  \label{fig:sim}
  \end{center}
\end{figure*}

\subsubsection{Simulation Results}
The results for the simulation are shown in \refFig{fig:sim}, where we investigate the influence of different hotspot CBS fractions $h$. For $h=0$ no hotspot CBSs exist and for $h=1$ all CBSs are hotspot CBSs. Solid lines are our new algorithm with the CBS prioritization step, dashed lines the algorithm from our previous work as a reference.

In Figures \ref{fig:simfeas0} to \ref{fig:simfeas1} we show the resulting feasibility of the CBS, i.e. the fraction of desired CBSs that were successfully established. Even for a scenario with no hotspots ($h=0$) the CBS prioritization step increases the CBS feasibility to $100\%$ for all capacity demands and all numbers of desired CBSs. In the scenarios with $h=0.5$ and $h=0.75$, the feasibility drops as the number of desired CBSs increases. Still the CBS prioritization step increases the CBS feasibility by over $20\%$. In the scenario with all desired CBSs in the hotspot ($h=1$), there is no significant improvement from prioritizing CBSs, which is the expected outcome. 

Figures \ref{fig:simwl0} to \ref{fig:simwl1} show the results for the overall number of used wavelengths and Figures \ref{fig:simwpl0} to \ref{fig:simwpl1} show the results for the number of used wavelengths per link. In the scenario with no hotspot CBSs ($h=0$) the CBS prioritization significantly improves the efficient use of wavelengths, as both the total number of used wavelengths and the number of wavelengths per link are significantly lower when using the CBS prioritization. This effect is also evident in the scenarios with $h=0.5$ and $h=0.75$. For $h=1$ there is again no significant difference. 

\subsection{Prototype Evaluation}
\label{sec:evalproto}
\begin{figure*}[tbh]
  \begin{center}
    \subfloat[A Proiri Feasibility]{
  \includegraphics[width=0.25\textwidth]{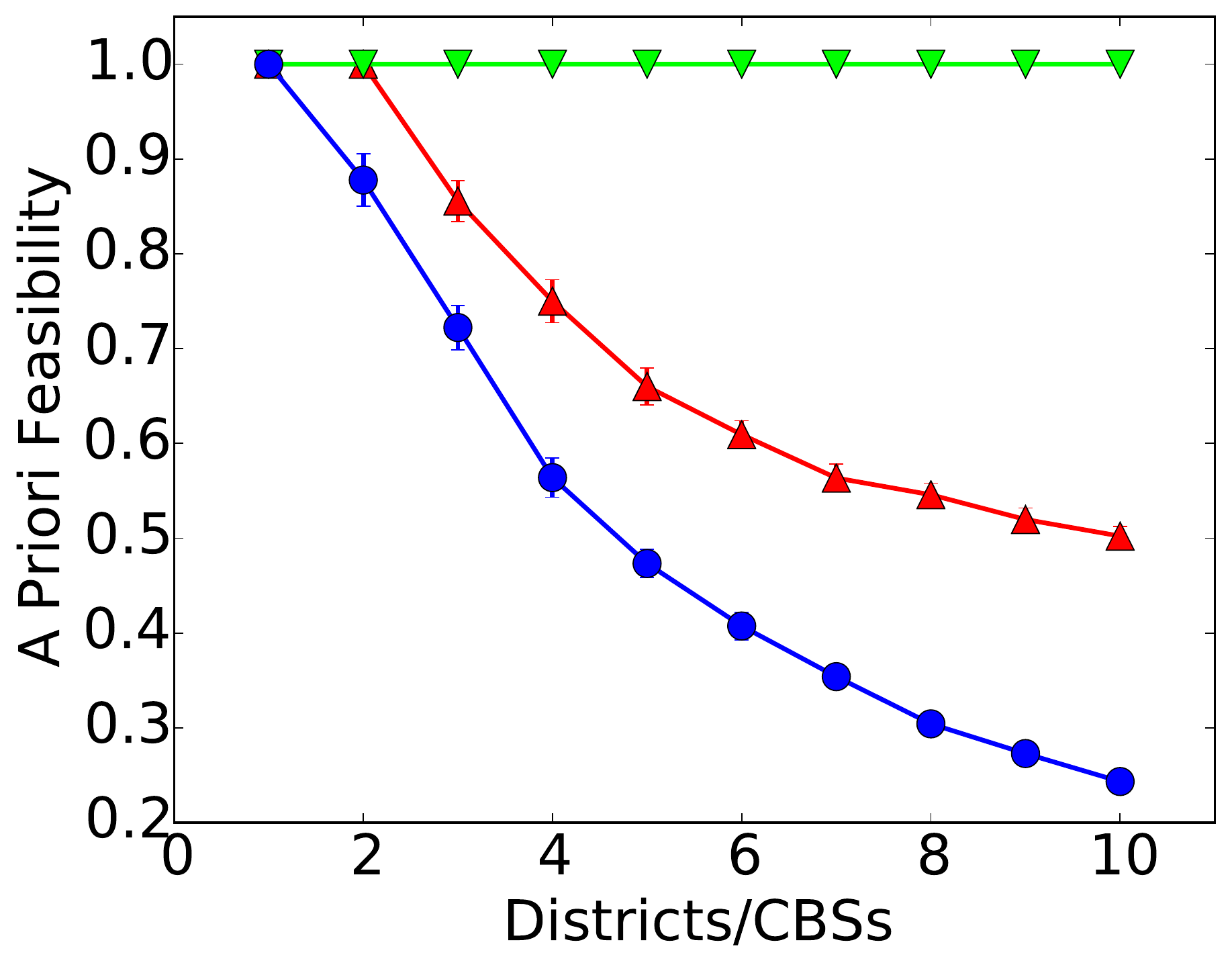}
  \label{fig:tbfeas}
  }%\\
    \subfloat[Achieved Throughput]{
  \includegraphics[width=0.25\textwidth]{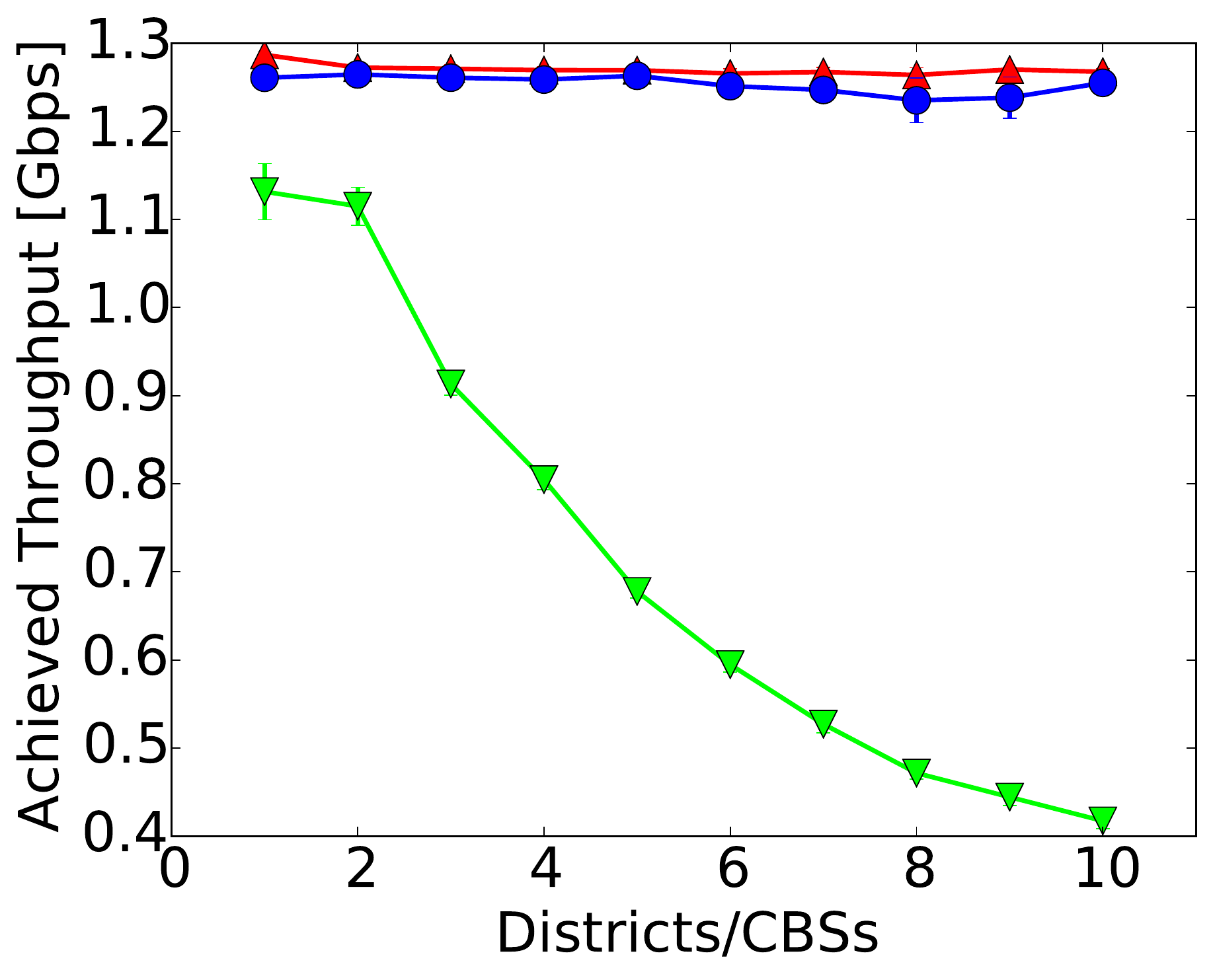}
  \label{fig:tbbw}
  }%\\
      \subfloat[Packet Loss]{
  \includegraphics[width=0.25\textwidth]{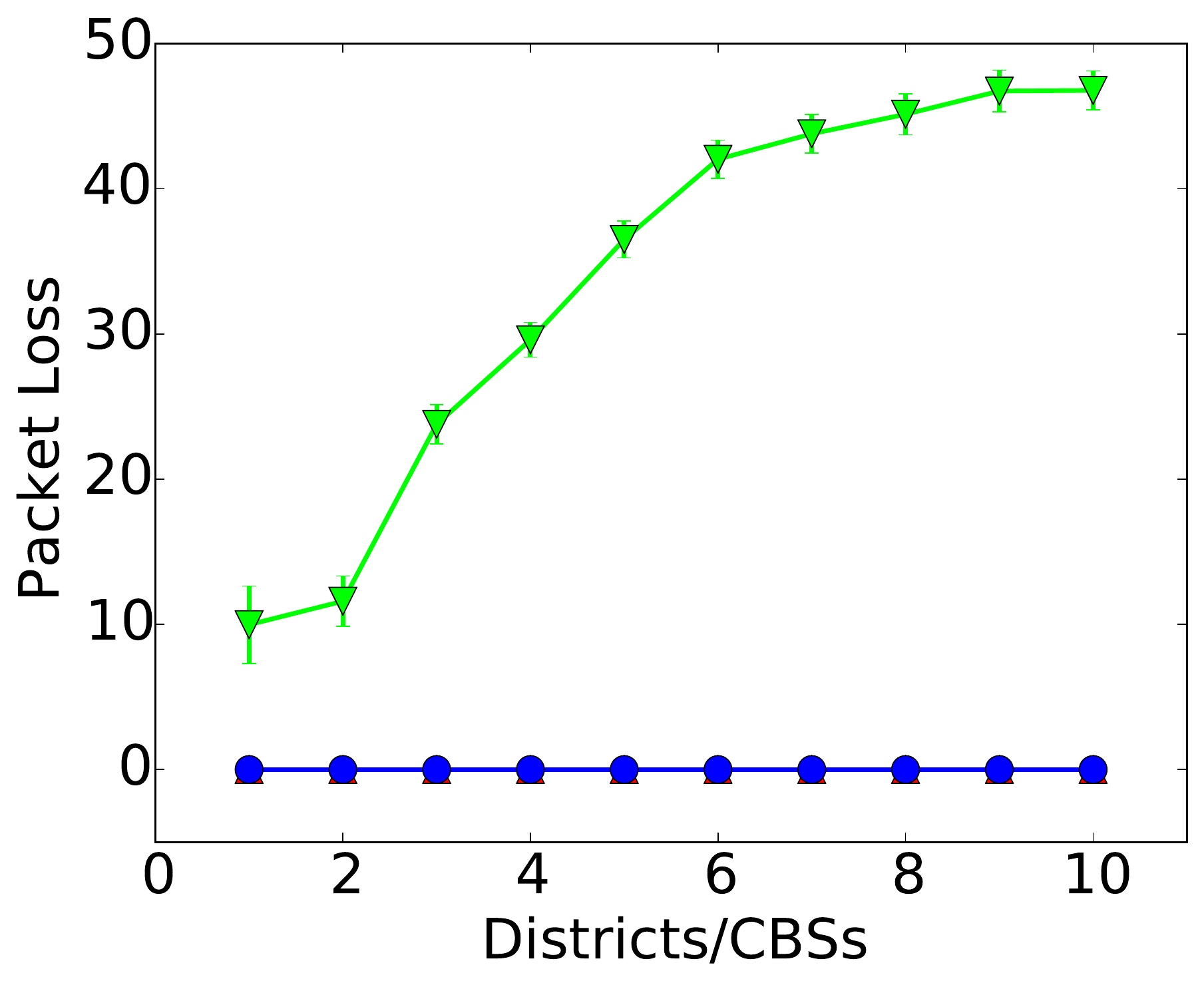}
  \label{fig:tbloss}
  }%\\
   \subfloat{
   \raisebox{1.4cm}{\includegraphics[width=0.2\textwidth]{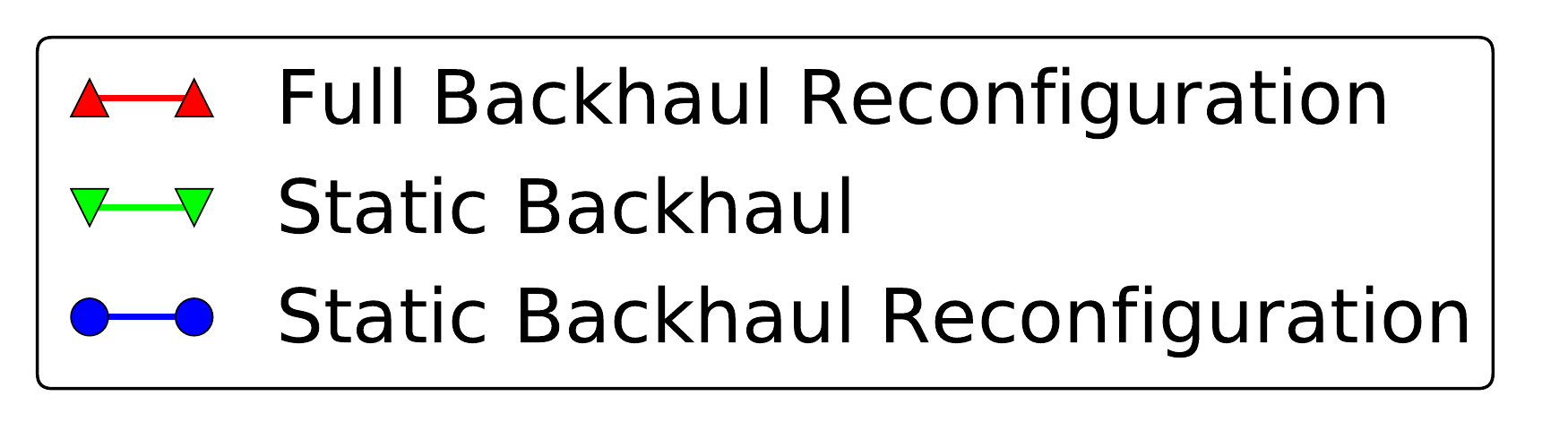}}
   \label{fig:tblegend}
   }
  \caption{Prototype evaluation}
  \label{fig:tb}
  \end{center}
\end{figure*}

The prototype evaluation uses our testbed implementation described in \refSec{sec:testbed}.

\subsubsection{Prototype Scenario}

As a scenario for the prototype evaluation we use a smaller scenario with only 16 BSs. To avoid a bottleneck from the Maxinet worker interconnect, a CBS can only contain BSs being emulated on the same worker machine. Apart from this limitation, CBSs are generated in the same random way as in the simulation (\refSec{sec:evalprio}). All emulated links also have a capacity of 2.5\,Gb/s, and for the demand per BS we only consider 1.25\,Gb/s, because the other values from the simulation (0.625\,Gb/s and 2.5\,Gb/s) do not provide additional insights from the evaluation results.

We also use three different implementations for the application to compare the performance of our algorithm:

\begin{itemize*}
\item \emph{Full Backhaul Reconfiguration} uses our full algorithm described in \refSec{sec:algo} with the full flexibility in terms of backhaul network configuration and BBU/LC placement.
\item \emph{Static Backhaul Reconfiguration} uses a limited version of our algorithm, where the BBU/LC is placed on a fixed BS and the algorithm only performs the flow routing and wavelength assignment.
\item \emph{Static Backhaul} does not use our algorithm at all and relies on a fixed BBU/LC placement, a static assignment of wavelengths and always uses shortest paths for the flow routing.
\end{itemize*}

\subsubsection{Prototype Results}

For the prototype evaluation, we consider three different metrics.

The \emph{a priori feasibility} is calculated from the output of the application. In \refFig{fig:tbfeas} we can see that the \emph{a priori feasibility} decreases for both applications that are based on our algorithm as the number of desired CBSs increases. This is due to the limited available resources in the backhaul network. The results also show that using the full flexibility of our algorithm yields a better \emph{a priori feasibility} than using the limited version. The \emph{a priori feasibility} for the static backhaul implementation is always 100\%, because this implementation does not consider any resource constraints in an a priori way.

In order to measure the \emph{a posteri feasibility} of all implementations we perform a UDP throughput measurement using iperf \cite{iperf} with a desired throughput of 1.25\,Gb/s. As results from this measurement we obtain both the achieved throughput (\refFig{fig:tbbw}) and the packet loss (\refFig{fig:tbloss}). We can see that both implementations with our algorithm achieve the desired throughput and do not cause any packet loss. In contrast to that the static implementation is not able to achieve the desired throughput and causes a significant packet loss. 
\section{Conclusion}
\label{sec:concl}

We have presented an extended approach to verify the feasibility of base station coordination in a centralized RAN environment considering capacity and latency constraints with a backhaul network with limited resources. This approach includes the dynamic assignment of backhaul resources, which we call backhaul network \emph{configuration}, as well as the dynamic instantiation of BBUs or local controllers (LCs).

Our simulation shows that our extension enables the use of our approach in \emph{dense} wireless access networks with hotspots of users, and furthermore increases the feasibility of base station coordination in networks without hotspots. This is a significant improvement compared to our previous work \cite{draexlerew2014}.

We have also presented a prototype implementation for a real-world deployment of our approach in a testbed. Our testbed measurements show that our approach is able to guarantee the availability of the desired resources for each feasible CBSs, and increase the feasibility of CBSs compared to a static assignment of backhaul resources. 

%\FloatBarrier

\section*{Acknowledgements}
The research leading to these results has received funding from the European Union’s Seventh Framework Programme (FP7/2007-2013) under grant agreement n$^\circ$ 318115.

\bibliography{bib}
\bibliographystyle{IEEEtran}

\end{document}